\documentclass[aps,prl,twocolumn]{revtex4}
\usepackage{makeidx}
\usepackage{CJK}
\usepackage{colordvi}
\usepackage{graphicx}
\usepackage{dcolumn}
\usepackage{bm,amsmath,verbatim}
\usepackage{mathrsfs}
\usepackage{color}
\usepackage[latin9]{inputenc}
\usepackage{amsmath}
\usepackage{amssymb}
\usepackage{graphicx}
\usepackage{esint}
\usepackage{amsfonts}
\usepackage{eurosym}
\usepackage{color}
\usepackage{xcolor}
\usepackage{mathrsfs}
\usepackage{amsmath}
\usepackage{graphicx}
\usepackage{dcolumn}
\usepackage{bm}
\usepackage{times}
\usepackage{amssymb}
\usepackage{float}
\usepackage{array}
\usepackage{tikz}
\DeclareMathOperator{\sech}{sech}
\usepackage[colorlinks,urlcolor=blue,citecolor=blue]{hyperref}

\begin{document}

\title{AC Oscillation of a Spin Soliton Driven by a Constant Force}
\author{Li-Chen Zhao$^{1,2}$}
\author{Wenlong Wang$^{3,4}$}
\author{Qinglin Tang$^{5}$}
\author{Zhan-Ying Yang$^{1,2}$}
\author{Wen-Li Yang$^{1,2}$}
\author{Jie Liu$^{6,7}$}
\email{jliu@gscaep.ac.cn}

\address{$^{1}$School of Physics, Northwest University, Xi'an 710127, China}
\address{$^{2}$Shaanxi Key Laboratory for Theoretical Physics Frontiers, Xi'an 710127, China}
\address{$^{3}$Department of Physics, Royal Institute of Technology, Stockholm, SE-106 91, Sweden}
\address{$^{4}$College of Physics, Sichuan University, Chengdu 610065, China}
\address{$^{5}$School of Mathematics, State Key Laboratory of Hydraulics and Mountain River Engineering, Sichuan University, Chengdu 610064, China}
\address{$^{6}$Graduate School, China Academy of Engineering Physics, Beijing 100193, China}
\address{$^{7}$CAPT, HEDPS, and IFSA Collaborative Innovation Center of the Ministry of Education, Peking University, Beijing 100871, China}

\begin{abstract}
\noindent$\displaystyle$The phenomena of AC oscillation generated by a DC drive, such as the famous Josephson AC effect in superconductors and Bloch oscillation in solid physics, are of great interest in physics. Here we report another example of such counter-intuitive phenomenon that a spin soliton in a two-component Bose-Einstein condensate is driven by a constant force: The initially static spin soliton first moves in a direction opposite to the force and then changes direction, showing an extraordinary AC oscillation in a long term. In sharp contrast to the Josephson AC effect and Bloch oscillation, we find that the nonlinear interactions play important roles and
the spin soliton can exhibit a periodic transition between negative and positive inertial mass even in the absence of periodic potentials. We then develop an explicit quasiparticle model that can account for this extraordinary oscillation satisfactorily. Important implications and possible applications of our finding are discussed.
\end{abstract}
\maketitle

\noindent$\displaystyle$The phenomena of AC oscillation generated by a DC drive are of great interest because of their counter-intuitive character  \cite{Barone,Makhlin,bloch,blochuse}. The Josephson AC effect is the  famous one. It was first predicted in the context of electron tunneling across an insulating barrier between two superconductors \cite{Jose}, in which a unidirectional driving voltage can result in oscillating electronic currents.
 The underlying mechanism is quantum phase coherence.
 The Bloch oscillation in solid physics is another example, which describes the motion of an electron in a periodic potential driven by a DC electric field  \cite{BO,BO2}.
 It is a direct consequence of the periodicity of the energy band structure that can induce a transition between the negative effective mass and positive mass\cite{Dahan}.
These striking phenomena not only are interesting in physics but also have important applications. For instance, a superconducting quantum interference device (SQUID) based on the Josephson effect has been invented that is extremely sensitive to magnetic measurements \cite{squid}.

In this paper, we report that a spin soliton in a cigar-shaped two-component Bose-Einstein condensate (BEC) can also demonstrate the AC oscillation in the presence of an external unidirectional constant force.
The oscillation frequency is  proportional to the force, and the
amplitude is inversely proportional to the force.
The underlying mechanism, however, is distinct from the phase-coherent mechanism of a typical Josephson oscillation in superconductors \cite{Jose} and many other Josephson-like oscillations in various quantum systems \cite{Liu,Levy,Polo,Cataliotti,Valtolina}.
We find that the inertial mass of the spin soliton can exhibit a periodic transition between negative and positive values because of a particular relation between its energy and moving velocity. This is somewhat similar to what occurs in the Bloch oscillation \cite{BO,BO2,Dahan}; however, the periodic potential is absent
and the nonlinear interactions between the degenerate atoms play an important role
in our situation. This inertial mass transition effect implies that the spin soliton can sometimes accelerate along the force direction and sometimes accelerate in the opposite direction, leading to an AC oscillation. We then develop a quasiparticle model to describe the motion of the spin soliton that can quantitatively account for this extraordinary oscillation. An experimental observation, important implications and a possible application to weak force measurements are discussed.
 \begin{figure*}[t!]
\begin{center}
\includegraphics[height=140mm,width=130mm]{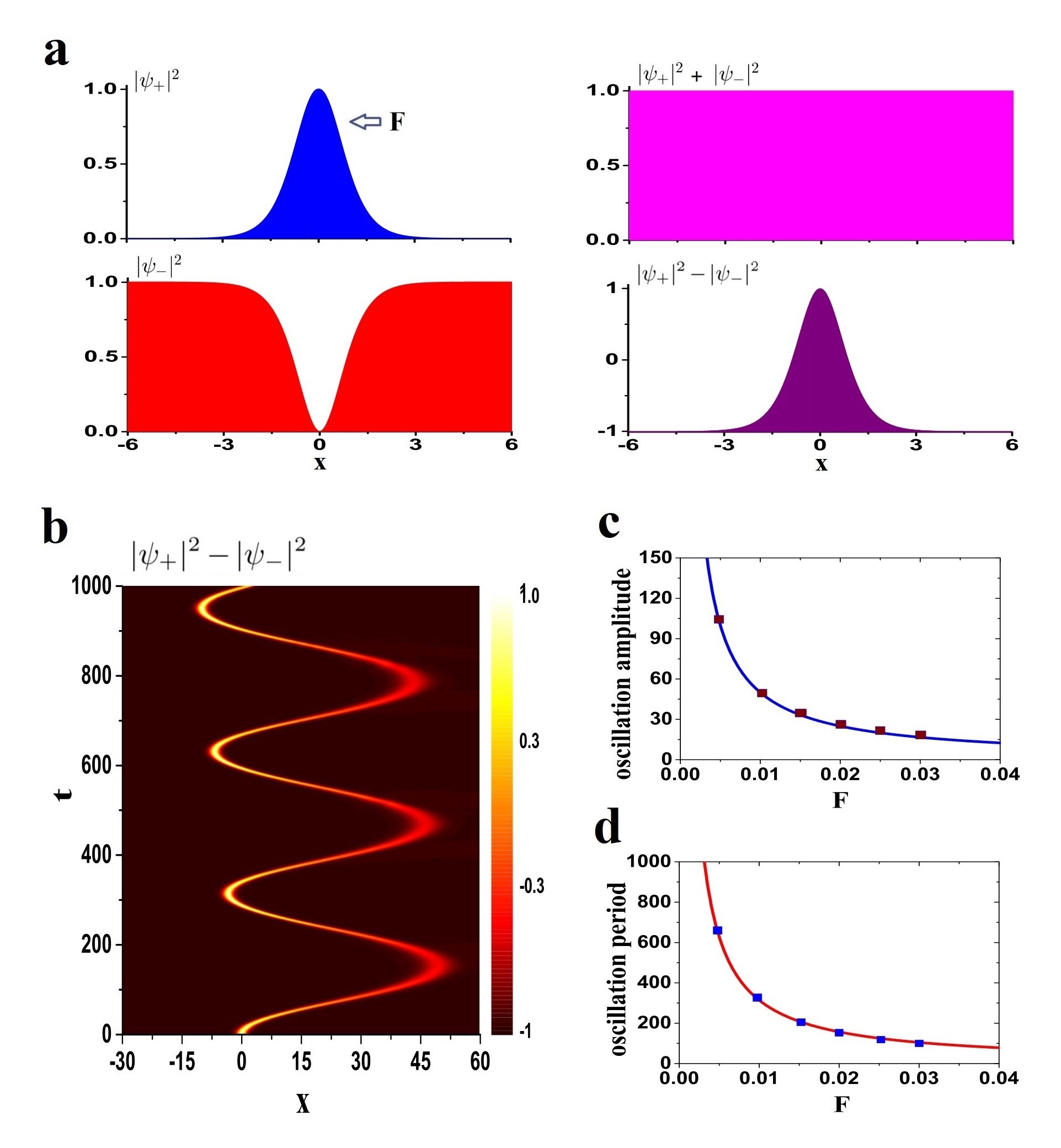}
\end{center}
\caption{ \textbf{ AC oscillation of spin soliton.} (\textbf{a}) Density profiles of the spin soliton. There is a bright soliton in the $\psi_+$ component and a dark soliton in the $\psi_-$ component. The whole particle density is uniform, and the pseudo-spin density distribution ($|\psi_+|^2-|\psi_-|^2$) admits a soliton. The parameters are $g_1=1$, $g_2=2$, $g_3=3$, and $v=0$. (\textbf{b}) Numerical evolution of the spin density with an external force of $F=-0.01$ adding on bright soliton component (the initial state  is given by (a)). The result shows that an AC oscillation emerges. (\textbf{c}) The oscillation amplitude vs. external constant force strength. (\textbf{d}) The oscillation period vs. force strength. The solid lines  are given by the analytical results $A = \frac{c_s^2} {2|F|}$ and $T = \frac{c_s \pi}{|F|}$. The square dots denote the numerical results.  }\label{Fig1}
\end{figure*}

\

\noindent$\displaystyle\textbf{Spin solitons in a two-component BEC}$. We consider a two-component BEC system which is tightly confined in the radial direction so that the radial characteristic length is smaller than the healing length and its dynamics is essentially one-dimensional (1D) \cite{Kevrekidis}. Rescaling the atomic mass and Planck's constant to be $1$, the mean-field energy for the quasi-1D BEC system can be written as $H= \int_{- \infty}^{+\infty} \psi_+^{*} (-\frac{1}{2} \partial^2_x) \psi_++\psi_-^{*} (-\frac{1}{2} \partial^2_x) \psi_-+\frac{g_1}{2} |\psi_+|^4+\frac{g_3}{2} |\psi_-|^4+g_2  |\psi_+|^2 |\psi_-|^2] d x$ \cite{Kevrekidis}. $x$ is the axial coordinate, and  $\psi=(\psi_+,\psi_-)^T $ denotes the condensate wave function, where $\pm $ refers to the two components. The dimensionless dynamical equations can be written as the following coupled model,
\begin{eqnarray}
&&i \frac{\partial \psi_+}{\partial t}= -\frac{1}{2} \frac{\partial^2 \psi_+}{\partial x^2} + (g_{1} |\psi_+|^2+g_{2} |\psi_-|^2) \psi_+, \\
&&i \frac{\partial \psi_-}{\partial t}= -\frac{1}{2} \frac{\partial^2 \psi_-}{\partial x^2} + (g_{2} |\psi_+|^2+g_{3} |\psi_-|^2) \psi_-.
\end{eqnarray}
The parameters $g_1$ and $g_3$ denote intraspecies interactions between the atoms in the components $\psi_+$ and $\psi_-$, respectively, and $g_2$ describes the interspecies interactions between the atoms.

For $g_1=g_2=g_3$, the system is described by an integrable Manakov model \cite{Mana}, and various types
of solitons have been deduced using the traditional inverse scattering method, B\"{a}cklund transformation
method and Hirota bilinear method \cite{Mat,Hirota,Zhao,Lingdnls}, such as bright-bright, bright-dark, and dark-dark solitons. Nevertheless, these solutions cannot be extended to non-Manakov cases where the constraint condition $g_1= g_2=g_3$ is not satisfied.
Here, we claim that under the conditions of $2g_2=g_1+g_3$ and $g_1\neq g_3$, we can derive an exact soliton solution with a the constraint condition $|\psi_{+}|^2+|\psi_-|^2 = C$ ($C=1$ for simplicity, see Appendix A for details). A soliton solution with $g_2>g_1$, as an example, can be written in the following explicit form:
\begin{eqnarray}
\psi_+(x,t)&=&\sqrt{\frac{c_s^2-v^2}{c_s^2}} \sech[\sqrt{c_s^2-v^2} (x- v t)] \nonumber\\&& e^{\frac{1}{2} i [-g_1  t-g_2  t+2 v (x-v t)]},
\end{eqnarray}
\begin{eqnarray}
\psi_-(x,t)&=&  \left(\sqrt{1-\frac{v^2}{c_s^2}} \tanh [\sqrt{c_s^2-v^2} (x- v t)]+\frac{i v}{c_s}\right)  \nonumber\\&&   e^{-i ( -g_1+2 g_2) t},
\end{eqnarray}
 where $c_s= \sqrt{g_2-g_1}$ denotes the maximum speed of soliton.
 The moving velocity $v$ of the soliton should be smaller than $c_s$, and when it equals the speed of sound, the above solution degenerates to a plane wave. When $v=0$, we have a static soliton, as shown in Fig. 1 (a). The particle density in the $\psi_+$ ($\psi_-$) component admits a bright soliton (dark soliton). The bright soliton in one component is induced by the effective potential generated by the dark soliton in the other component \cite{Carr}.

 The total density distribution of the soliton solution is uniform, i.e., $|\psi_{+}|^2+|\psi_-|^2 = 1$, while the population imbalance  of $|\psi_+|^2-|\psi_-|^2$ ranged in $[-1,1]$ exhibits an explicit soliton profile. This differs from that of the usual dark-bright solitons reported previously, where the sum density distribution also shows a soliton profile \cite{Busch,Nistazakis,Carr,BEC,Hamner}. For each position, the solution has the analogous structures of a spin-half system represented by a Bloch sphere \cite{pspin,Liu2,Liu3,Liu4,Liu5}, therefore, we term it as spin soliton.
A linear stability analysis has indicated that the spin soliton is stable (see Appendix B).
Note that the spin soliton emerges in the phase separation regime with $g_2^2>g_1 g_3$, the background densities of its two components are different and equal to zero and unit, respectively.
In contrast, for  the ``magnetic soliton" found very recently \cite{Qu},
the background densities of its two components have the same value of $0.5$ with a unit total density background. It has the character of phase miscibility and emerges in the regime of $g_2^2<g_1 g_3$.

\

\noindent$\displaystyle\textbf{The AC oscillation of a spin soliton}$. We now investigate the dynamics of the spin soliton driven by a constant force.
 Initially, the spin soliton is set to be static, as shown in Fig. 1 (a). A weak unidirectional force (sketched by Fig. 1 (a)) or, equivalently, a linear potential $-F x$, is added only to the bright soliton component $\psi_+$ to avoid accelerating the whole particle density background. In simulations, a term of $\int_{- \infty}^{+\infty} -F x |\psi_+|^2 dx$  is added to the  mean-field energy. Here, ``weak" means that the external potential varies slowly over the size scale of the soliton \cite{Busch}; therefore, it cannot destroy the soliton structure. We solve the nonlinear Schr\"odinger equation numerically in a spatial range of $[-600,600]$ by the discrete Cosine transform method with homogeneous Neumann boundary conditions \cite{Yang,Tang}.

We chose $F=-0.01$ to demonstrate our results. Strikingly, the spin soliton moves in a direction opposite to the force for a while and then changes direction, showing an oscillation over the long term, as shown by the spin density evolution in Fig. 1 (b). During the evolution, the whole particle density remains almost uniform with only an approximately $5\%$ mass density fluctuation.
We perform further numerical calculations to investigate the dependence of the oscillation amplitude $A$ and period $T$ on external force $F$. The results are shown in Fig. 1 (c) and (d), respectively.
It is clearly shown that the oscillation frequency is proportional to the force and the amplitude is inversely proportional to the oscillation frequency.

Our extensive simulations show that the oscillation behavior can emerge even for a bit higher or lower bright soliton component, when the force strength is smaller than $0.05$.
We have tuned the coupling strengths by $\pm 0.1$ in absolute magnitudes deviated from  $g_1=1, g_2=2$ and $g_3=3$, the AC oscillation still can be observed clearly. It indicates that the striking oscillation phenomenon  is rather robust.

\begin{figure}[t!]
\begin{center}
\includegraphics[height=40mm,width=85mm]{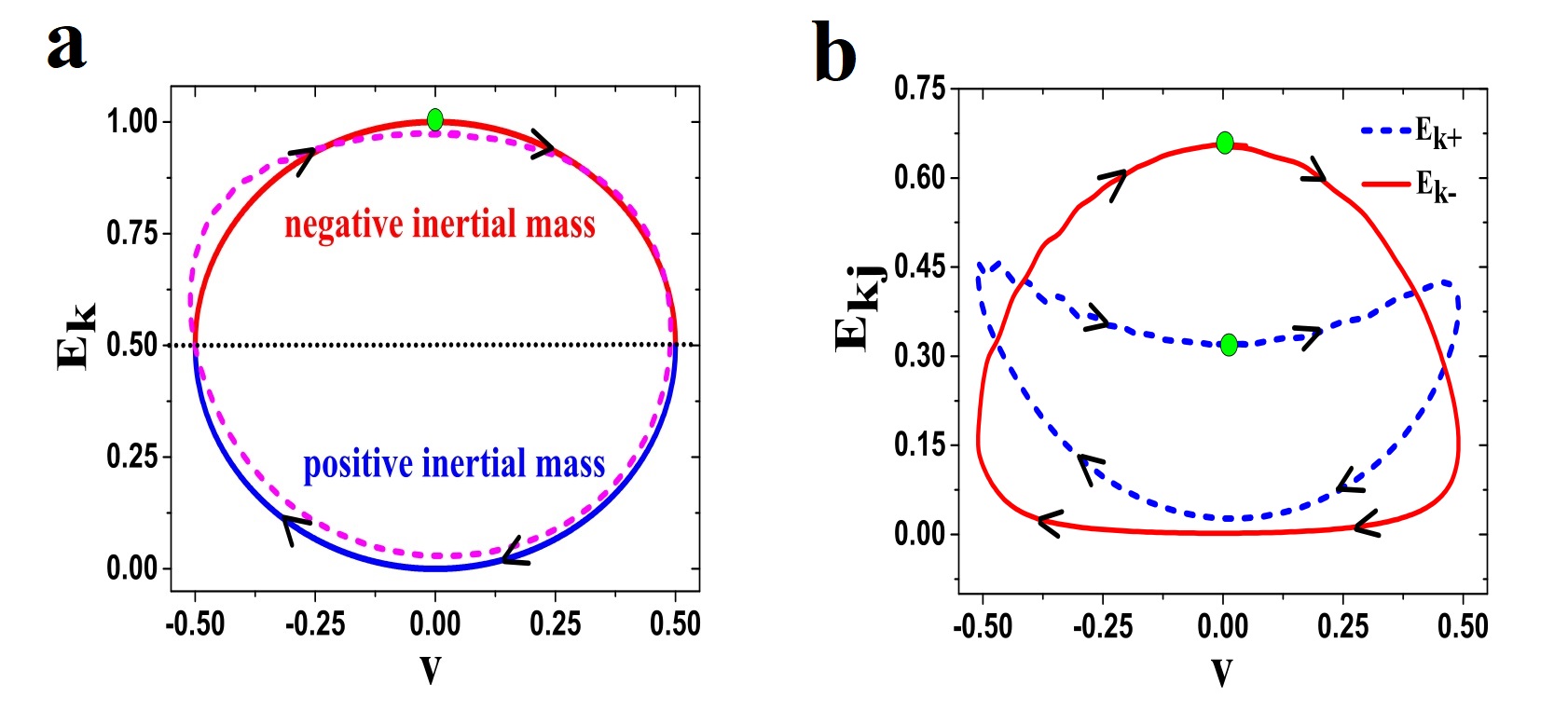}
\end{center}
\caption{ \textbf{The mechanism of AC oscillation.} (\textbf{a}) The relation between the kinetic energy of the spin soliton and its moving velocity. The purple dashed line denotes numerical results, and the solid line is given by $(E_k-c_s/2)^2+v^2 = (c_s/2)^2$ derived from Lagrangian variational method. The spin soliton admits both negative mass (upper semicircle) and positive mass (lower semicircle) during one oscillation period. (\textbf{b}) The numerical relations between the kinetic energy of the soliton in each BEC component and its velocity. The bright soliton admits positive mass (blue dashed line) and the dark soliton mainly has negative mass, except near the maximum velocities (red solid line). The competition between them enables the spin soliton to admit both positive and negative mass. The green dot denotes the initial state for the AC oscillation in Fig. 1 (b). The black arrows indicate the evolution direction. The parameters $g_1=1$, $g_2=2$, $g_3=3$, $F=-0.01$. }\label{Fig3}
\end{figure}

\

\noindent$\displaystyle\textbf{Negative-positive mass transition}$. To understand this striking oscillation behavior, we first investigate the kinetic energy of the spin soliton.
The exact spin soliton solution of the explicit expressions (3-4) cannot describe the acceleration process because, in the presence of an external force, the spin soliton will evolve with a broadening or shrinking of its width and change shape.
Thus, we have to calculate the kinetic energy of the spin soliton (the interaction energy keeps nearly zero for a spin soliton) by directly solving the nonlinear Schr\"odinger equation according to $E_k=\int_{-L_1}^{+L_2} \psi_+^{*} (-\frac{1}{2} \partial^2_x) \psi_++\psi_-^{*} (-\frac{1}{2} \partial^2_x) \psi_- dx$. The parameter $L_j$ is chosen to be a bit larger than the soliton size, i.e., $L_1=30$ and $L_2=80$. Our extensive numerical calculations suggest a simple approximate relation between the kinetic energy and moving velocity of $(E_k-c_s/2)^2+v^2 = (c_s/2)^2$ \cite{note1}, which gives two branches of $E_k = c_s/2\pm \sqrt{c_s^2/4-v^2}$, as shown in Fig. 2 (a). This explicit dispersion relation is further verified analytically
according to the Lagrangian variational method (see Appendix C for details).

The density profile of the spin soliton is spatially localized during the whole evolution (see Fig. 1 (b)); therefore, the spin soliton can be viewed as a quasiparticle. The inertial mass of the spin soliton can be derived from the relation between the soliton energy $E_s$ and velocity according to $M^*=2  \frac{\partial E_s} {\partial (v^2)}=2  \frac{\partial E_k} {\partial (v^2)}$ \cite{Brand}, i.e.,
\begin{eqnarray}
M^*&=&\mp \frac{2/c_s} {\sqrt{1-v^2/(c_s/2)^2}}.
\end{eqnarray}
The inertial mass of the spin soliton is shown in Fig. 2 (a).
It is seen that the spin soliton admits both negative mass (upper semicircle) and positive mass (lower semicircle) during each oscillation period.

We also calculate the inertial mass for the dark soliton and bright soliton separately according to the individual kinetic energy in each component of the BEC. The results are shown in Fig. 2 (b). The relations also agrees well with the ones given by Lagrangian method. We see that the bright soliton admits positive inertial mass, and the dark soliton admits mainly negative mass, similar to scalar soliton systems \cite{Lewenstein,PNAS}. However,
in contrast to a scalar soliton, the density profile of the bright soliton obtained here depends on the moving velocity, and its inertial mass varies with the velocity accordingly (see the blue dashed line in Fig. 2(b)). The dark soliton (red solid line in Fig. 2(b)), however, might exhibit positive mass around the maximum velocities.

When applying an external force, the bright soliton initially tends to move along the direction of the force. At same time, it drags the dark soliton to move along the force direction because the interaction between the dark soliton and bright soliton is indeed attractive due to the repulsive interaction between the two components. However, the dark soliton admits a relatively larger negative mass, implying that it prefers to move against the drag force (i.e., buoyancy effect) and can dominate the initial motion direction of the spin soliton. In the following temporal evolution, due to the interplay between the bright and dark solitons, the total inertial mass of the spin soliton can periodically change from negative to positive values.  In contrast to the oscillation behavior of the magnetic soliton that is directly induced by the axial harmonic trap potential \cite{Qu}, the striking oscillation of our spin soliton emerges in the absence of axial trapping potential and is due to the intrinsic mechanism of positive-negative mass transition.

Negative mass is an interesting subject \cite{Bondi,Kromer,Bonnor} and is even believed to play an important role in the expansion of the early universe. Negative mass has also been reported in BEC systems. Recently, an experimental observation of negative mass effects was realized through the engineering of the dispersion relation by spin-orbit coupling effects \cite{negativemass}, in which negative mass leads to dynamical instability and
 a sudden increase in the atomic density. Recently, interactions of solitons with positive and negative masses were investigated in a BEC trapped by an optical-lattice potential \cite{Sakaguchi}.

\

\noindent$\displaystyle\textbf{Quasiparticle model}$. The concept of the inertial mass captures the response of the spin soliton to an applied force, encapsulating Newton's equations of quasiparticle dynamics. The external potential energy of a soliton $E_p=\int_{- \infty}^{+\infty} -F x |\psi_+|^2 dx = -2 F x_c/c_s $ ($x_c$ denotes the soliton center position). The force acting on the spin soliton is then $-\frac{dE_p}{dx_c}=2F/c_s$. Thus, the dynamical trajectory of the spin soliton should be governed by $2F/c_s=M^* \frac{d^2 x_c}{dt^2}$.  Let us consider the explicit expression of the inertial mass (5) and set the initial conditions to $t=0, x_c=v=0$. The analytical solution of the above Newton equation is readily obtained as follows:
\begin{equation}
x_c=- \frac{c_s^2}{2F} \sin^2(F t/c_s).
 \end{equation}
We can see that the oscillation amplitude $A = \frac{c_s^2} {2|F|}$ and period  $T = \frac{c_s \pi}{|F|}$, which agrees well with numerical simulations (see Fig. 1 (c) and (d)).
 \begin{figure}[t!]
\begin{center}
\includegraphics[height=90mm,width=80mm]{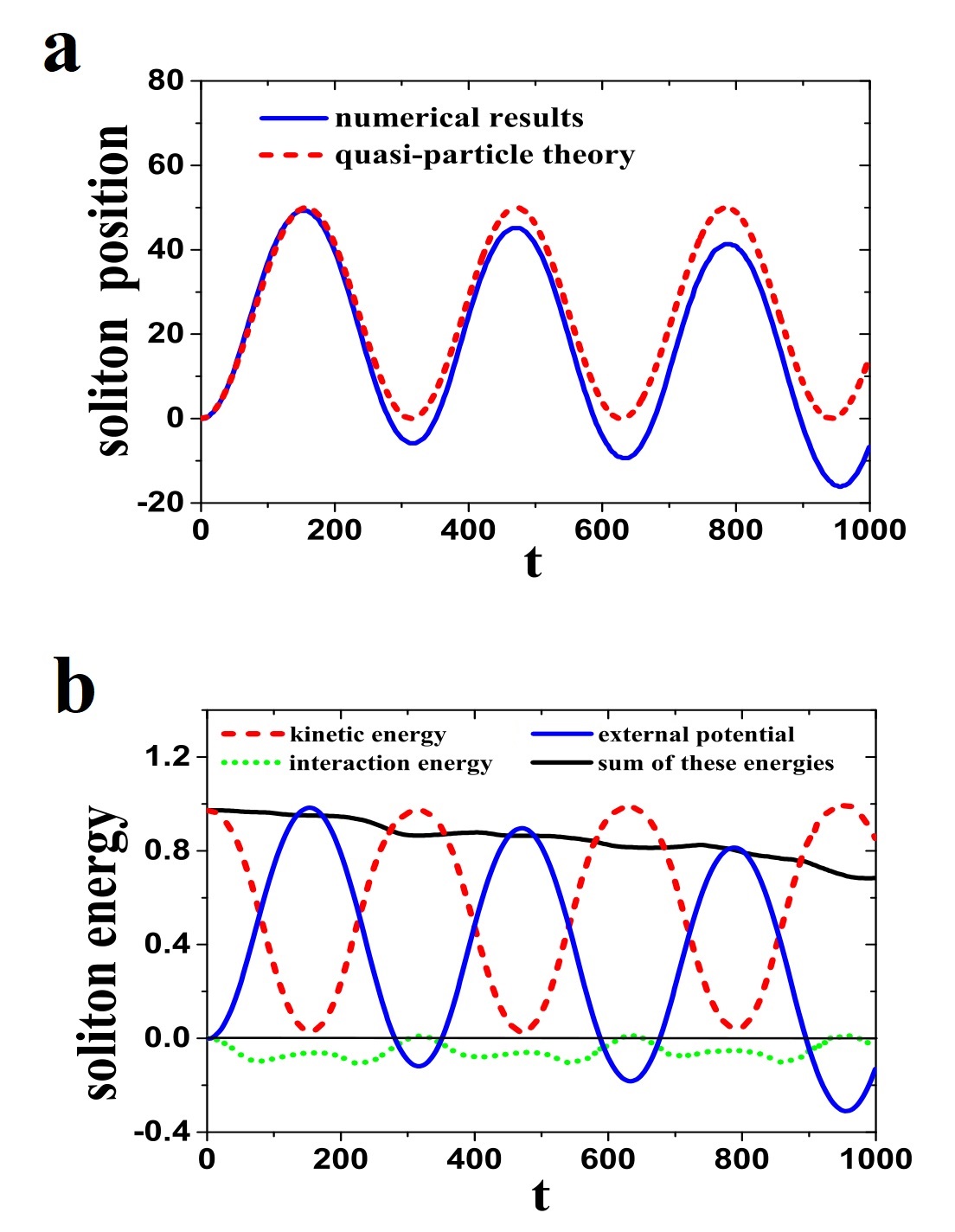}
\end{center}
\caption{ \textbf{Quasiparticle theory description.} (\textbf{a}) A comparison of spin soliton trajectories from quasiparticle theory and numerical simulations for the AC oscillation in Fig. 1 (b). The periodic oscillation can be well predicted by quasiparticle theory except for a small downward shift in the trajectory induced by the dissipation of soliton energy, as discussed in the text. (\textbf{b}) The corresponding temporal evolution of the kinetic energy, external potential energy, interaction energy, and sum of these energies.
Compared to the other two types of energy, the interaction energy remains small. The small decay in the energy sum and the downward shift in the external potential energy are due to the soliton energy spreading to other regimes through the excitation of dispersive waves or other nonlinear waves. For details, see the text. The parameters $g_1=1$, $g_2=2$, $g_3=3$, $F=-0.01$. }\label{Fig4}
\end{figure}
 We have compared the theoretical prediction with numerical simulations for the AC oscillation in Fig. 1 (b). As shown in Fig. 3 (a), both the oscillation amplitude and period can be well predicted by the simple model, except for a small downward shift in the trajectory.
To understand this deviation, in Fig. 3 (b), we integrate over the local soliton profile (safely in $[-30,80]$ regime) and plot the temporal evolution of the kinetic energy $E_k$, external potential energy $E_p$, interaction energy of soliton $E_{inter}=\int_{- \infty}^{+\infty} [\frac{g_1}{2} |\psi_+|^4+\frac{g_3}{2} (|\psi_-|^2-1)^2+g_2  |\psi_+|^2 (|\psi_-|^2-1)] dx$ \cite{Carr,Kivshar}, and sum of these energies. Compared to the other two types of energy, the total interaction energy $E_{inter}$ sum over three terms keeps nearly zero, while each of them is not. Note that the interaction plays an important role manifesting that the dispersion relation or inertial mass are explicitly dependent on the interaction parameters. The kinetic energy $E_k$ oscillates periodically, and there is a periodic transition between the kinetic energy and external potential energy. However,
the external potential energy shifts downward, and the energy sum shows an
 ``unphysical" decay. These effects are due to the soliton energy spreading to other regimes through the excitation of dispersive waves or other nonlinear waves. Our numerical simulations indicate that dispersive waves mainly emerge in the dark soliton component and are almost absent in the bright soliton component. The total energy in the full space $[-600,600]$  is conserved in our numerical simulations.
Due to total energy conservation, with an increase in the dispersive wave energy, the external potential energy of the soliton will decrease, leading to a deviation in the spin soliton trajectory from our quasiparticle model.
\begin{figure}
\includegraphics[height=55mm,width=85mm]{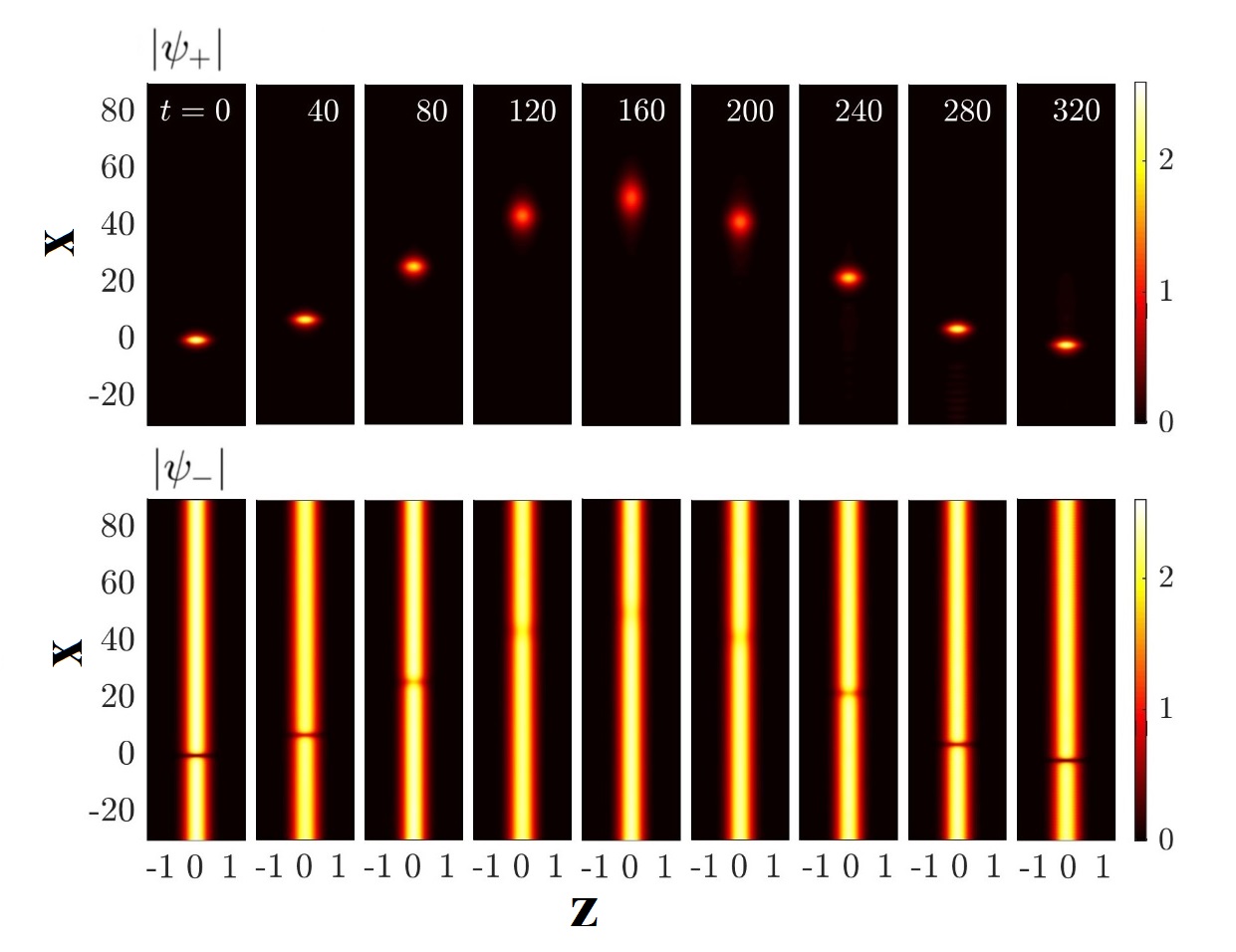}
\caption{\textbf{The spin soliton AC oscillation dynamics in a 3D setting}.  The dynamics agrees well with the effective 1D counterpart, cf. Fig. 1. One cycle is illustrated. Note that the condensate has rotational symmetry along the $x$-axis and hence only the $x$-$z$ cross section is shown. The parameters $\omega_x=0$, $\omega_{\perp}=20$, $g_1=1$, $g_2=2$, $g_3=3$, $F=-0.01$.
}
\label{fig4}
\end{figure}

\

\noindent$\displaystyle\textbf{The AC oscillation in 3D setting}$. We now extend to investigate the AC oscillation in three-dimensional (3D) setting, that is, the BECs are trapped by a harmonic trap $\frac{1}{2} \omega_{\bot}^2 (y^2+z^2)+\frac{1}{2}\omega_{x}^2 x^2$. Here, we use a strongly confining transverse frequency $\omega_{\perp}=20$ to ensure that radial characteristic length  is smaller than the healing length for quasi-1D approximation (radial characteristic length is $0.22$, and the healing length is about $0.71$ in this case). We will show that the striking AC oscillation emerges and is rather robust in the genuinely 3D situation. The 3D dynamical equations and simulation methods are given Appendix D.

We first set $\omega_x=0$ corresponding to the above quasi-1D BEC, the results presented in Fig. 4 demonstrate a perfect one cycle oscillation (see the movie \cite{dbmv} for a more complete dynamics to $t=1000$). The oscillation period ($\approx 320$) agrees to its 1D counterpart as well as our theory.
\begin{figure}
\includegraphics[height=55mm,width=85mm]{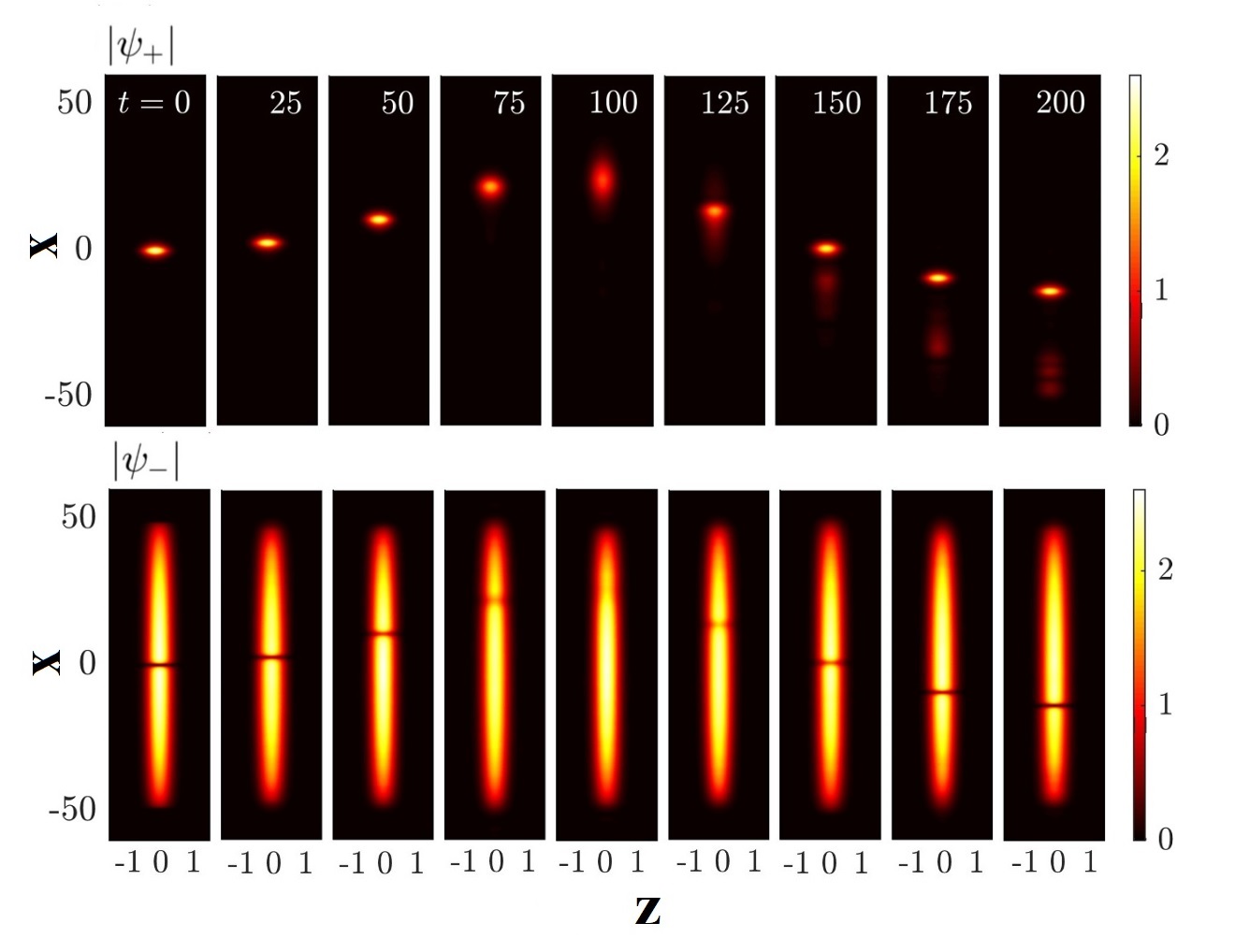}
\caption{\textbf{The spin soliton oscillation dynamics in a harmonic trap.} One can still observe a robust oscillation of spin soliton.  One cycle is illustrated.   The oscillation period (amplitude) is much shorter (smaller) than the ones in Fig. 4. The condensate has rotational symmetry along the $x$-axis and hence only the $x$-$z$ cross section is shown. The parameters $\omega_x=0.05$, $\omega_{\perp}=20$, $g_1=1$, $g_2=2$, $g_3=3$, $F=-0.01$.
}
\label{fig5}
\end{figure}

When $\omega_x= 0.05$, i.e., a harmonic trap along the $x$-axis presents, we have still observed a kind of periodic oscillation shown in Fig. 5 (see the movie \cite{dbmv2} for a more complete dynamics to $t=1000$). Its period is about $200$, which is however much shorter than the one in Fig. 4. In this situation, the intrinsic AC oscillation has been dramatically influenced by the external potential. One can see some smear of the bright soliton component in Fig. 5, due to the nonlinear excitation generated  by the external harmonic trap. Therefore, to observe our intrinsic AC oscillation, the  external trap
should be weak enough satisfying the limitation $\omega_x \ll  \frac{2 |F|}{c_s}$.

\

\noindent$\displaystyle\textbf{Discussion}$.
We demonstrate that AC oscillation emerges for a driven spin soliton in a two-component BEC and reveal its distinctive mechanism associated with the negative-positive mass transition. Our numerical simulations indicate that spin soliton and the AC oscillation are stable and robust and  can emerge in genuinely 3D situation. This striking phenomenon is expected to be observed in current experiments.

Let us consider ultra-cold $^{87}Rb$ atoms prepared in the internal states $\left\vert F = 1, m_F = -1\right\rangle $ and $\left\vert F = 2, m_F = 0\right\rangle $ (denoted by $\psi_+$ and $\psi_-$, respectively). For hyperfine states, the scattering lengths can be manipulated by external magnetic fields \cite{Widera,Tojo,Mertes}, which can be used to ensure that the nonlinear interaction strength nearly satisfies the condition $2g_2= g_1+g_3$ for spin solitons. Recent experiments indicated that vector solitons can be prepared well in BEC systems \cite{BEC,Hamner,Bersano}. Our numerical simulation also indicates that the AC oscillation phenomenon of a spin soliton is robust against a low level of noise and some
parameter deviations from ideal condition.  A weak magnetic field can be applied along the principal axis of the cigar-shaped BEC to drive the bright soliton in the $\left\vert F = 1, m_F = -1\right\rangle $ state without influencing the $\left\vert F = 2, m_F = 0\right\rangle $ component.

In principle, the AC oscillation phenomenon of a spin soliton can be used to diagnose weak forces or related physical quantities through a direct measurement of the moving period of ultracold atoms, for instance, the cigar-shaped BEC with a spin soliton could serve as a bubble level instrument that can work in a microgravity environment. This approach offers an alternative to the approach employed in recent experiments with optomechanical systems \cite{Schreppler,Moller}, where forces are determined through the measurement of optical frequencies.

\

\

\noindent$\displaystyle\textbf{Acknowledgments}$

\noindent$\displaystyle$Stimulating discussions with Xi-Wang Luo, and Jun-Peng Cao are acknowledged. Zhao is grateful to Ling-Zheng Meng and Peng Gao for their help in numerical simulations. This work is supported by the National Natural Science Foundation of China (Contract No. 11775176, 11775030, 11674034, U1930403), the Major Basic Research Program of Natural Science of Shaanxi Province (Grant No. 2018KJXX-094), and the Key Innovative Research Team of Quantum Many-Body Theory and Quantum Control in Shaanxi Province (Grant No. 2017KCT-12).  W.W.~acknowledges support from the Swedish Research
Council Grant No.~642-2013-7837, and the Goran Gustafsson Foundation
for Research in Natural Sciences and Medicine, and the Fundamental Research Funds for the Central Universities, China.

\

\

\noindent$\displaystyle\textbf{Appendix A:}$ The method of deriving exact soliton solution

\noindent$\displaystyle$To obtain spin solitons, we firstly set a constrain condition on the mass density distributions $|\psi_{+}|^2+|\psi_-|^2 = 1$. With this condition, we can further simplify the Eq. (1) and (2) as follows,
\begin{eqnarray}
&&i \frac{\partial \psi_+}{\partial t}+ \frac{1}{2} \frac{\partial^2 \psi_+}{\partial x^2} +(g_{2}-g_{1}) |\psi_+|^2 \psi_+- g_{2}  \psi_+=0,\nonumber\\
&&i \frac{\partial \psi_-}{\partial t}+ \frac{1}{2} \frac{\partial^2 \psi_-}{\partial x^2} +(g_{2}-g_{3}) |\psi_-|^2 \psi_--g_{2} \psi_-=0.
\end{eqnarray}
If $g_2-g_1$ and $g_2-g_3$ are both negative or positive, there are dark solitons or bright solitons in the two components. Obviously, the superposition of them can not be unform at all. Therefore, we need the second constrain condition that $g_2-g_1$ and $g_2-g_3$ have different signs  for spin solitons.
In this case, there are one dark soliton and one bright soliton in the two components respectively, and it is possible to satisfy the condition $|\psi_{+}|^2+|\psi_-|^2 = 1$. We choose $g_2-g_1 >0$ and $g_2-g_3 <0$ to derive spin solitons analytically and exactly,  from the well-known results of scalar nonlinear Schr$\ddot{o}$dinger equation. Then, we can give static bright soliton and dark soliton solution of Eq. (7) as follows
\begin{eqnarray}
\psi_+&=& \frac{\sqrt{f_1}} {\sqrt{g_2-g_1}}   \sech[\sqrt{f_1} x] \ e^{i  f_1/2 t-i g_2 t}, \nonumber \\
\psi_-&=&\frac{\sqrt{f_2}} {\sqrt{g_3-g_2}}  \tanh[\sqrt{f_2} x]  e^{-i f_2 t-i g_2 t},
\end{eqnarray}
where $f_1$ and $f_2$ determine the amplitude of bright soliton and plane wave background for dark soliton component respectively. Finally, the constrain condition $|\psi_{+}|^2+|\psi_-|^2 = 1$ further gives that $f_1=f_2=g_2-g_1$, and $g_1-g_2=g_2-g_3$. In this way, we construct a static spin soliton solution  of Eq. (1) and (2) as follows
\begin{eqnarray}
\psi_+(x,t)&=& \sech[c_s x]  e^{\frac{1}{2} i [-g_1  t-g_2  t]}, \nonumber\\
\psi_-(x,t)&=&  \tanh [c_s x]   e^{-i ( -g_1+2 g_2) t},
\end{eqnarray}
 where $c_s= \sqrt{g_2-g_1}$ denotes the speed of sound. One can derive spin soliton solution with velocity in similar ways, and the spin soliton solution is given in the text.
This means that it is possible to construct exact spin soliton solutions with the condition $g_1+g_3=2g_2$. It should be noted that the spin soliton solutions fails to hold for the case $g_1=g_2=g_3$, for which the coupled model becomes the well-known integrable Manakov model, and mass solitons exist with many different forms.

 \begin{figure}[t!]
\begin{center}
\includegraphics[height=60mm,width=80mm]{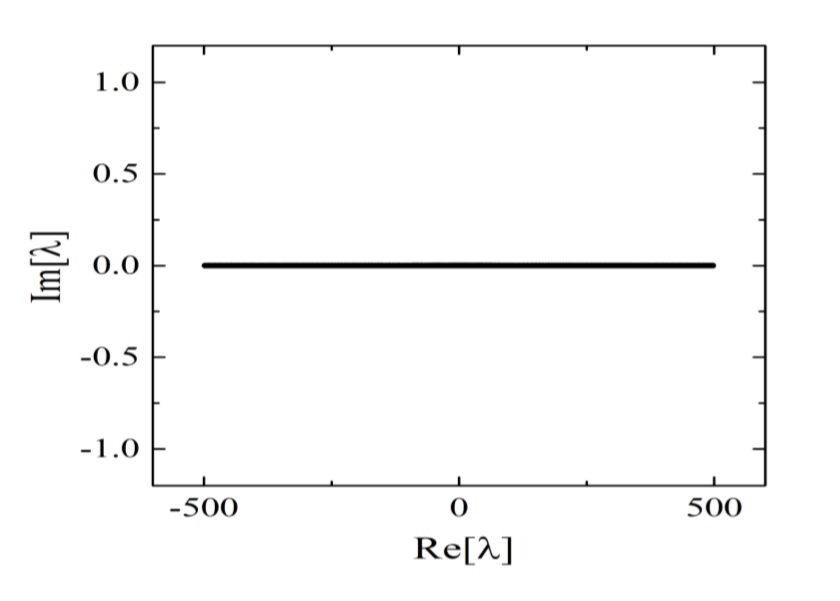}
\end{center}
\caption{The excitation spectrum of the spin soliton. It is seen that spin soliton admits spectral stability. }\label{FigS1}
\end{figure}

\

\noindent$\displaystyle\textbf{Appendix B:}$ The stability of spin soliton in 1D

\noindent$\displaystyle$We perform linear stability analysis on the spin soliton. Introducing weak perturbations on the spin soliton, $\psi_{+p}=\psi_+ (1+P_+(x) e^{i \lambda t}+Q_+(x) e^{-i\lambda^* t}) $, $\psi_{-p}=\psi_- (1+P_-(x) e^{i \lambda t}+Q_-(x) e^{-i\lambda^* t})$ (where $\psi_+$ and $\psi_-$ are the spin soliton solution), we can obtain linearized equation for the eigenvalue of $\lambda$. The excitation spectrum is shown in Fig. 6. $Im[\lambda]=0$ indicates that spin soliton is stable. Numerical simulations also indicate that spin soliton is indeed robust against noises in one dimension case.

\

\noindent$\displaystyle\textbf{Appendix C:}$ The analytic derivation of the relation between soliton energy and velocity by Lagrangian variational method

\noindent$\displaystyle$The dynamical equation of a two-component BEC driven by a constant force $F$ can be written as follows,
\begin{eqnarray}
&&i \frac{\partial \psi_+}{\partial t}= -\frac{1}{2} \frac{\partial^2 \psi_+}{\partial x^2} + (g_{1} |\psi_+|^2+g_{2} |\psi_-|^2) \psi_+-Fx \psi_+,\nonumber\\
&&i \frac{\partial \psi_-}{\partial t}= -\frac{1}{2} \frac{\partial^2 \psi_-}{\partial x^2} + (g_{2} |\psi_+|^2+g_{3} |\psi_-|^2) \psi_-.
\end{eqnarray}
With considering  the  constrain condition $|\psi_{+}|^2+|\psi_-|^2 = 1$, i.e., we can simplify Eq.~(10) as follows,
\begin{eqnarray}
&&i \frac{\partial \psi_+}{\partial t}+ \frac{1}{2} \frac{\partial^2 \psi_+}{\partial x^2} + c_s^2 |\psi_+|^2 \psi_+- g_{2}  \psi_++F x \psi_+=0,\nonumber\\
&&i \frac{\partial \psi_-}{\partial t}+ \frac{1}{2} \frac{\partial^2 \psi_-}{\partial x^2} - c_s^2 |\psi_-|^2 \psi_--g_{2} \psi_-=0,
\end{eqnarray}
where $c_s^2=g_2-g_1=g_3-g_2$.
The above equation can be further written as:
\begin{eqnarray}
&&i \frac{\partial \psi_+}{\partial T}+ \frac{1}{2} \frac{\partial^2 \psi_+}{\partial X^2} +  |\psi_+|^2 \psi_+- g_{2}/c_s^2  \psi_++\frac{F X}{c_s^3} \psi_+=0,\nonumber\\
&&i \frac{\partial \psi_-}{\partial T}+ \frac{1}{2} \frac{\partial^2 \psi_-}{\partial X^2} - |\psi_-|^2 \psi_--g_{2}/c_s^2 \psi_-=0,
\end{eqnarray}
where $X=c_s x$, $T=c_s^2 t$ which are introduced to simplify the following calculations.

In the presence of the force $F$ term, the system become non-integrable and the exact analytic expressions for the
dynamic evolution of the spin solitons can not be obtained. We thus exploit the Lagrangian variational method to evaluate the dynamics of the spin soliton by introducing the following trial functions,
\begin{eqnarray}
\psi_+&=& f(T) \sech[(X-b(T))/w(T)]\nonumber\\
  && e^{i \phi_0(T)+i\phi_1(T) (X-b(T))}, \nonumber \\
\psi_-&=&\{i \sqrt{1-f(T)^2}+f(T) \tanh[(X-b(T))/w(T)]\}\nonumber\\
&& e^{i \theta_0(T)}.
\end{eqnarray}
Note that the total density keeps a constant in temporal evolution. The soliton position, amplitudes and width vary in time.

We now  use the Lagrangian variational method to derive expressions of $b(T),f(T), w(T), \phi_1(T), \phi_0(T), \theta_0(T)$.
The Lagrangian of the system is: $L(t)=\int_{-\infty}^{+\infty} [\frac{i}{2} (\psi_+^* \partial_t \psi_+-\psi_+ \partial_t \psi_+^*)+\frac{i}{2} (\psi_-^* \partial_t \psi_--\psi_- \partial_t \psi_-^*) (1-\frac{1}{|\psi_-|^2})-\frac{1}{2} |\partial_x \psi_+|^2-\frac{1}{2} |\partial_x \psi_-|^2-\frac{g_1}{2} |\psi_+|^4-\frac{g_3}{2} (|\psi_-|^2-1)^2-g_2|\psi_+|^2 (|\psi_-|^2-1)+F x |\psi_+|^2] dx=\int_{-\infty}^{+\infty} \{c_s [\frac{i}{2} (\psi_+^* \partial_T \psi_+-\psi_+ \partial_T \psi_+^*)+\frac{i}{2} (\psi_-^* \partial_T \psi_--\psi_- \partial_T \psi_-^*) (1-\frac{1}{|\psi_-|^2})-\frac{1}{2} |\partial_X \psi_+|^2-\frac{1}{2} |\partial_X \psi_-|^2]-\frac{1}{c_s} [\frac{g_1}{2} |\psi_+|^4+\frac{g_3}{2} (|\psi_-|^2-1)^2+g_2|\psi_+|^2 (|\psi_-|^2-1)]+\frac{1}{c_s^2} F X |\psi_+|^2]\} dX=L(T)$.
The factor $(1-\frac{1}{|\psi_-|^2})$ is introduced for the dark soliton state, or it is impossible to integrate the term $+\frac{i}{2} (\psi_-^* \partial_t \psi_--\psi_- \partial_t \psi_-^*)$. This problem was firstly solved by Yuri S. Kivshar \textit{et al.} in 1995 \cite{Kivshar}. It should be noted that $g_2-g_1=g_3-g_2$ is kept for spin soliton. Substituting the trial wavefunctions into the Lagrangian, and after taking the particularly elaborate integrals, we obtain that
$L(T)= c_s \{2 f(T)^2 w(T) (\phi_1(T) b'(T) -\phi'_0(T))-\frac{f(T)^2}{w(T)} (1+\phi_1(T)^2 w(T)^2)+ 2 f(T)^2 w(T) \theta'_0 +2 (\arcsin[f(T)]-f(T) \sqrt{1-f(T)^2}) b'(T)\} + \frac{1} {c_s^2} 2 F f(T)^2 w(T) b(T)$,
where $b'(T)=\frac{d}{d T}b(T)$, etc. Our initial conditions are $f(0)=w(0)=1$, $b(0)=b'(0)=0$. From the conservation of the norm of the bright component, we have $w(T)=1/f(T)^2$. It is now straightforward to apply the Lagrangian equation $\frac{d}{dT} (\frac{\partial L(T)} {\partial \alpha'})=\frac{\partial L(T)}{\partial \alpha}$, where $\alpha=b(T),f(T), \phi_1(T), \phi_0(T), \theta_0(T)$. We have obtained three nontrivial equations along with two trivial equations, the three nontrivial equations are  $(\arcsin[f])' = \frac{F}{c_s^3}$,
$b' =f \sqrt{1-f^2}$,
$\phi_1 = b'$.
Using the initial conditions, we find the following solution:
\begin{eqnarray}
f(T) &=& \cos(FT/c_s^3),\nonumber\\
 b(T) &=& \pm \frac{c_s^3} {2F} \sin^2(FT/c_s^3),\nonumber\\
\phi_1(T) &=& b'(T)=\pm \cos(FT/c_s^3) \sin(FT/c_s^3).
\end{eqnarray}

Substituting the trial functions into kinetic energy $E_k=\int_{-\infty}^{+\infty} \psi_+^{*} (-\frac{1}{2} \partial^2_x) \psi_++\psi_-^{*} (-\frac{1}{2} \partial^2_x) \psi_- dx$, we obtain the dispersion relation
\begin{eqnarray}
E_k&=&\frac{c_s} {2} \pm \sqrt{(\frac{c_s}{2})^2-v^2}.
\end{eqnarray}
This also means that $(E_k-c_s/2)^2+v^2 = (c_s/2)^2$, which agrees well with the numerical simulation results in main text. The interaction energy can be also calculated $E_{inter}= \int_{- \infty}^{+\infty} [\frac{g_1}{2} |\psi_+|^4+\frac{g_3}{2} (|\psi_-|^2-1)^2+g_2  |\psi_+|^2 (|\psi_-|^2-1)] dx=0$.

Moreover, the soliton center and speed evolve as: $x_c = X_c/c_s=b(T)/c_s=\pm \frac{c_s^2}{2F} \sin^2(Ft/c_s)$, $v=\frac{dx_c}{dt}= \pm \frac{c_s}{2} \sin(2F t/c_s)$.
 This indicates that the spin soliton oscillates periodically in the presence of a constant force.
  These results  agree perfectly with the results calculated from the quasi-particle model in the main text.

\

\noindent$\displaystyle\textbf{Appendix D:}$ The 3D dynamical equations for AC oscillation

\noindent$\displaystyle$The evolution of spin soliton in 3D case in the presence of a harmonic trap can be described by
\begin{eqnarray}
      &&i \frac{\partial\psi_{+}}{\partial t}+\frac{1}{2}\Delta \psi_{+}-(g_{1}^{3D}|\psi_{+}|^2+g_{2}^{3D}|\psi_{-}|^2)\psi_{+}\nonumber\\
      &&-[\omega_{\bot}^2 (y^2+z^2)/2+\omega_{x}^2 x^2/2-F x]\psi_{+} = 0, \nonumber \\
     && i \frac{\partial\psi_{-}}{\partial t}+\frac{1}{2}\Delta \psi_{-}-(g_{2}^{3D}|\psi_{+}|^2+g_{3}^{3D}|\psi_{-}|^2)\psi_{-}\nonumber\\
     &&-[\omega_{\bot}^2 (y^2+z^2)/2+\omega_{x}^2 x^2/2]\psi_{-} = 0,
\end{eqnarray}
where $g_{j}^{3D} =\frac{2\pi}{\omega_{\bot}}g_{j}$. Other parameters are the same as those of the homogeneous case.
In order to perverse the feature of our spin soliton, we study the Thomas-Fermi regime i.e. we use a weak harmonic trap of frequency $\omega_x$. In this case, the dark soliton is modulated by the Thomas-Fermi ground state $\sqrt{\max(1-\omega_x^2 x^2/2,0)}$. The initial states are $\psi_{+}= \sech[c_{s}x]\ \sqrt{\frac{\omega_{\bot}}{\pi}}e^{-\frac{1}{2}\omega_{\bot}(y^2+z^2)}$,  $\psi_{-}=\sqrt{\max(1-\omega_x^2 x^2/2,0)} \tanh[c_{s}x]\ \sqrt{\frac{\omega_{\bot}}{\pi}} e^{-\frac{1}{2}\omega_{\bot}(y^2+z^2)}$.
With $\omega_x=0$, $\omega_{\perp}=20$, and $F=-0.01$, we integrate the Gross-Pitaevskii equation numerically using the finite element method and the fourth-order Runge-Kutta method. We have applied periodic boundary conditions to the bright soliton component and homogeneous Neumann boundary conditions to the dark soliton component. With $\omega_x=0.05$, $\omega_{\perp}=20$, and $F=-0.01$, we solve the Gross-Pitaevskii equation numerically using the finite element method and the fourth-order Runge-Kutta method, and we applied periodic boundary conditions to both the bright and dark components as the condensate is trapped.

\

\

%
%


\begin{thebibliography}{99}
\bibitem{note1} This  relation is independent of both the initial position and weak forces, but it varies with the initial velocity of the spin soliton. For instance, when the initial velocity is set to be $0.2c_s$, the maximum velocity is found to shift from $c_s/2$ to 0.6 $c_s$.
\end{thebibliography}
\end{document}